\def\ni{\noindent }
\def\ie{{\it i.\ e.\ }}
\documentclass[12pt]{iopart}
\usepackage{iopams}  
\usepackage{epsfig}
\begin{document}

\title{Sphere  rolling  on the surface of a cone}

\author{I Campos,\dag J L Fern\'andez-Chapou,\ddag\ A L Salas-Brito,*\footnote[2]{On sabbatical leave from Laboratorio de Sistemas Din\'amicos, UAM-Azcapotzalco, Mexico City D~F, Mexico, email: asb@correo.azc.uam.mx}   C A Vargas,\ddag \footnote{Corresponding author. E-mail: cvargas@correo.azc.uam.mx} }

\address{\dag\  Facultad de Ciencias, Universidad Nacional Aut\'onoma de M\'exico,
Apartado Postal 21-939,  Mexico City 04000 D F, M\'exico}

\address{\ddag\ Departamento de Ciencias B\'asicas, Universidad Aut\'onoma Metropolitana, Unidad Azcapotzalco, Apartado Postal 21-267 Coyoacan, Mexico City  04000 D F, M\'exico}

\address{* Nonlinear Dynamical Systems Group, Department of Mathematics and Statistics, San Diego State University, 5500 Campanile Drive, San Diego, CA 92182-7720, USA
}

\begin{abstract}
We analyse the motion of a sphere that rolls without slipping on a conical surface having its axis in the direction of the constant gravitational field of the Earth. This nonholonomic system admits 
a solution in terms of quadratures. We exhibit that  the only circular of the system orbit is stable and furthermore show that all its solutions can be found  using an analogy with central force problems. We also discuss the case of motion with no gravitational field, that is, of motion on a freely falling cone.  \end{abstract}


\submitto{\EJP}

\maketitle

\section{Introduction}

Rigid body motion has always been an interesting and  very usable subject of classical mechanics \cite{ll,sommerfeld}. Many subtle points of dynamics and of  mathematical techniques for studying the behaviour of physical systems can be  learned by studying the motion of rigid bodies. On this matter see, for example, \cite{goldstein, flores72, soodak02, lopez02, 96, theron02,  carnero97, previous}. Part of the interest of today  comes from the insight that can be gained on the behaviour of spinning asteroids or artificial satellites and, furthermore,  rigid body motion  can be chaotic \cite{efroimsky00, macie03, ranada95}. Besides, many everyday phenomena can  be understood  in terms of rigid body motion at least in a first approximation. For example, balls rolling on inclines, motion in toboggans, bowls in a bowling alley, the motion of snow boarders, the dynamics of bicycles or   wheels, the behaviour of billiard balls, and  so on.   Though the importance of rigid bodies is clear, some of the problems involving  rolling particles on a surface are often modelled in beginning courses as point particles sliding on  surfaces \cite{lopez02,fajans00}. This modelling is an appropriate pedagogical device in  introductory courses but we want to show here that  the problems can be addressed using  a  rigid body approach in more advanced courses. 

 In this work  we analyse  the motion  of a sphere that rolls without slipping on the inside of a right circular cone under the influence of a uniform gravitational field  acting verically downwards, in the direction of the symmetry axis of the cone.  The motion of a  spherical body rolling without slipping on surfaces of revolution has been recently studied  with the purpose of illuminating control processes \cite{previous}. Here our aim is to study the motion of a sphere on the inner surface of a conical surfaceas an exactly solvable example of rigid body motion. We obtain the general solution of this problem expressing it in quadratures. We analyse certain qualitative features of the motion, like the existence of a stable circular orbit,  establishing an analogy   with particle motion in a central force field in two dimensions. For the case of a conical surface in free fall we find that the general solution of the problem can be casted in terms of  expressions  similar to those defining sets of ellipses and hyperbolas.  We  also calculate the apsidal (apogee) angle of the center of mass (CM) motion to  find the orbit's symmetry axes and argue that the sphere's CM trajectory densely fills a strip on the conical surface.

\section{The equations of motion}
To describe the position of the sphere's centre of mass (CM), we choose a Cartesian coordinate system $(x,y,z)$ such that the origin of coordinates is at the position of the sphere's centre of mass (CM) when the sphere simply rests on the cone (that is, it corresponds to the vertex of the imaginary cone ---shown dashed in figure 1--- on which the CM moves); the coordinates are then (see Figure 1)

\begin{equation}\label{5}
\eqalign{x&=r\sin\alpha \cos\vartheta,\cr
y&=r\sin\alpha\sin\vartheta,\cr
z&=r\cos\alpha,}
\end{equation}

\ni where $\alpha$ is  half the angle of aperture of the cone ---therefore, $0\le \alpha \le \pi/2$, $\alpha=\pi/2$ corresponds to motion on a plane, whereas $\alpha=0$ to motion on a cylindrical surface---,  $r$ is the distance from the origin to the CM, and $\vartheta$ is the polar angle of the sphere's CM.  All these relations may be seen from figure 1. The tangential components of the CM velocity, in the, respectively, meridional and parallel directions to the conical surface (using a sort of geographical terminology),    are

\begin{equation}\label{6}
\eqalign{ u&=\dot{r},\cr
v&=r\sin\alpha\dot{\vartheta}.}
\end{equation}

\begin{figure}[tbp] 
\centerline{ 
\epsfig{file=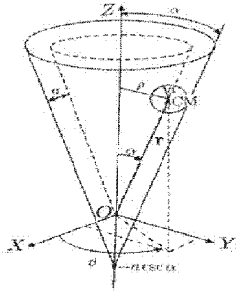, width=7cm,height=7cm} 
} 
\caption{The cone on which the sphere of radius $a$ rolls is shown in a continuous black line. The 
dashed cone is 
the  imaginary cone on which the sphere's centre of mass (CM) moves. The imaginary cone 
is the same as the 
actual one but displaced upwards a distance $a \csc \alpha $. The origin of 
coordinates is chosen as the vertex of the imaginary cone so the vertex of the actual cone is at 
$z_0=-a \csc \alpha$. The figure also shows the CM generalised coordinates $r$ and 
$\vartheta$. The radius of the parallel circle containing the CM is symbolised by $\rho$.   
} 
\label{fig1} 
\end{figure} 

\noindent Therefore,  Lagrangian of the system  can be written as

\begin{equation}\label{L1}
L=\frac {1} {2}M (u^2+v^2) + \frac {1} {2}  I [ \dot\phi^2 +2 \dot\phi \dot \psi \cos \theta +\dot \theta^2 + \dot\phi^2]- M g z
\end{equation}

\noindent where $M$ is the mass, $I=5Ma^2/2$,  is the moment of inertia of the sphere,  $a$ its radius; $g$ is the acceleration of gravity, and $\phi$, $\theta$ and $\psi$ are  Euler's angles \cite{ll}. Lagrangian (\ref{L1})  with $u$, $v$, and $z$ defined appropriately, is fairly general; the description it gives is  valid for an arbitrary surface of revolution with a sphere rolling on its surface.

The motion of the sphere is further constrained according to
$u - a{\mathbf \omega}\cdot \hat{\bf e}_1=0$, and $v+a {\mathbf \omega}\cdot \hat{\bf e}_2=0$, where ${\mathbf \omega}$ is the angular velocity vector of the sphere, and $\hat{\mathbf e}_1= (\sin\alpha
 \cos\vartheta,\; \sin\alpha\sin\vartheta, \cos\alpha)$, $\hat{\mathbf e}_2=(-\sin\vartheta, \cos\vartheta, 0)$ are unit vectors in the, respectively, direction of the meridional line and the parallel circle  on the cone. The vector ${\mathbf \omega}$ can be written in terms of the Euler angles, an expression that  can be found in many textbooks \cite{ll,sommerfeld,goldstein}. If we employ equations (\ref{6}) for substituting $u$ and $v$, we obtain
\begin{eqnarray}\label{constricciones} 
\frac{\dot r} {a}- \dot \theta \sin(\phi-\vartheta)+\dot\psi \sin\theta\cos(\phi-\vartheta)&=&0,\cr\cr
  \dot\vartheta\sin\alpha \,\frac{r}{a}+\dot\theta\sin\alpha\cos(\phi-\vartheta)-\dot\phi\sin\alpha + \cr
\dot\psi \left[ \sin\alpha\sin\theta \sin(\phi-\vartheta)+\cos\alpha\cos\theta \right]&=&0;
\end{eqnarray}
\noindent these two relations just mean that the sphere rolls with no slipping on the surface of the cone. 

The equations of motion of the sphere  can be obtained, from $L$ and  constrictions (\ref{constricciones}),  using Lagrange's  undetermined coefficient method\cite{sommerfeld}, to get 

\begin{eqnarray}\label{1}
\dot{u}+v\left[{v\over \rho} \sin \alpha  -\frac{2}{7}a\kappa \omega\right] =-{5\over 7}g\cos \alpha,\cr
\dot{v}-u\left[{v\over\rho}\sin\alpha\right]=0,\cr 
a\dot{\omega}+ uv \kappa=0,
\end{eqnarray}

\noindent  where $\rho$ is the radius of curvature of the parallel circle at point of contact: $\rho= r\sin\alpha$, $\kappa=\cos\alpha/\rho$,  and $\omega\equiv \omega_3$ is the component of ${\mathbf \omega}$ normal to the cone. 

The first two  equations above are essentially Newton's second law for the centre of mass of the sphere,  the third one is its  rotating counterpart for a rotating rigid sphere. With the help of the constrictions, the Euler's angles have been  eliminated from the description.

\section{Conserved Quantities in the Motion of the Sphere}

It should be clear that the total energy, $E$, of the sphere is a constant of motion since we are neglecting any dissipative effects. The energy of the sphere can be obtained directly from equations (\ref{1}), 

\begin{equation}\label{2}
E={7\over 10}M(u^2 + v^2)+{7\over 10}I\omega^2+ {7\over 5}Mgz.
\end{equation}

\noindent  We note in passing that this expression (\ref{2}) is valid, again with $u$ and $v$ properly defined, for any surface of revolution. The point $z=0$ has been chosen as the point of zero potential energy \cite{ll,previous}.  It can be worth noting that the factor $7/10$ in the CM's kinetic energy,  instead of the  usual $1/2$, comes from effects associated with  the rolling-but-no-slipping condition.

For arbitrary surfaces of revolution, the energy is the only constant of motion.  But in our conical case  it is simple to show, using the equations of motion (\ref{1}),  that  the $z$-component of the angular momentum of the center of mass (CM) is also a constant of the motion, 

\begin{equation}\label{3}
L_z=M\rho v=M\rho^2\dot{\vartheta}.
\end{equation}

\noindent Moreover, for spheres rolling on flat surfaces ---with symmetry axis as $z$-axis--- there exist one more constant of motion given by \cite{previous} 

\begin{equation}\label{4}
L_+=l-{2a b^2  \over 5(z-c)}L_z
\end{equation}

\ni with $b$ and $c$ constants coming from the parametrization,  in cylindrical coordinates, of such surfaces as 

\begin{equation}\label{3b}
z=b\rho+c,
\end{equation}

\noindent where $l=I\omega$ is the angular momentum of the sphere respect to the normal to the surface of revolution at the point of contact and $\omega$ the rotational angular speed around this same axis.  
In the conical surface $b$ is the slope of the generatrix  and $c$ is the $z$-coordinate of its vertex.
The easiest way of showing that (\ref{4}) is a constant is by  taking its $t$-derivative and using (\ref{1}) to show that it vanishes. In is also worth pinpointing that $z\ge 0\ge z_{0}\equiv -a\csc \alpha $  since the sphere is geometrically constrained---\ie\ we are assuming that it is not possible for the sphere to cut through the surface of the cone; $z_{0}$ being the $z$-coordinate of the vertex of the cone on which the sphere is rolling.

\noindent Using equation (\ref{6}) and the expression for $\rho$ in the integrals of motion (\ref{2}), (\ref{3}) and (\ref{4}), we obtain

\begin{equation}\label{7}
{1 \over 2} M (\dot{r}^2+ r^2 \dot{\vartheta}^2 \sin^2 \alpha  )+
{1\over 7} M a^2 \omega^2 + {5\over 7}Mgr\cos\alpha = {5\over 7}E,
\end{equation}

\begin{equation}\label{8}
Mr^2\dot{\vartheta}\sin^2\alpha=L_{z},
\end{equation}

\begin{equation}\label{9}
a\omega-{L_{z} \cos \alpha \over Mr\sin^2\alpha}= {2\,L_+\over 5\, M a}\equiv h,
\end{equation}

\noindent where $h$ is a new constant. Notice that equation (\ref{8}) shows that $\vartheta$ never changes sign, \ie\ the change in $\vartheta$ is monotonic. 

\section{The equivalent radial problem}

Combining  the previous equations, we can get a new expression for $r$ which can be expressed as

\begin{equation}\label{10}
{\cal E}={1\over 2}M\dot{r}^2+{A\over r^2}+{B\over r}+Cr,
\end{equation}

\noindent with ${\cal E}$, $A$, $B$ and $C$ constants defined by

\begin{equation}\label{11}
\eqalign{ 
{\cal E}&={5\over 7}E-{1\over 7}Mh^2,\cr
A&={L_{z}^2(1+{2\over 7}\cot^2\alpha)\over 2M\sin^2\alpha},\cr
B&={2\over 7}{L_{z}h\over \sin\alpha}\cot\alpha,\cr
C&={5\over 7}Mg\cos\alpha,\cr}
\end{equation}

\noindent in the case the sphere is rotating around the cone axis [$L_z\neq 0$ and $0< \alpha < \pi/2$] both  $A$ and $C$ are  strictly positive constants.  

\begin{figure}[tbp] 
\centerline{ 
\epsfig{file=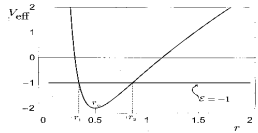, width=7cm,height=7cm} 
} 
\caption{The effective potential $V_{\hbox{eff}}(r)$ is plotted against $r$ in arbitrary units.   
A particular value of the constant `energy' ${\cal E}=-1$  is also shown. 
The points marked $r_1$ and $r_2$ are turning points, that is, points in
 which $\dot r =0$. As the plot exhibits the effective potential has  only 
one minimum and thus just two turning points for any ${\cal E}> V_{\hbox{eff}}(r_c) $. The plot also exhibits  that  $V_{\hbox{eff}}(r) $ attains its 
minimum value at $r_c$ and that this is  a stable minimum. The plot 
was 
made using $A=1$, $B=-3.5$, $C=2$ in equation (15).   
} 
\label{fig2} 
\end{figure} 

Expression (\ref{10}) can be regarded as the energy equation for an equivalent, purely `radial' problem, with an effective potential energy term given by

\begin{equation}\label{vef}
V_{\hbox{eff}}(r)={A\over r^2}+{B\over r}+Cr,
\end{equation}

\noindent which we show in figure 2. In this figure we have also traced a constant   ${\cal E} (=-1, \hbox{ in arbitrary units})$ value as the horizontal line crossing $V_{\hbox{eff}}$ in two points, $r_1$ and $r_2$---the turning points where $\dot r$ vanishes. This figure shows the typical behaviour of $V_{\hbox{eff}}$. A glance to figure 2 thus shows that the motion is restricted between the two turning points,  meaning that the sphere's CM  moves on a strip bounded by two parallel circles on the conical surface. 

The graph of $V_{\hbox{eff}}$, figure 2,  also shows a minimum value at $r_c$. That is, an {\sl stable  circular orbit} is possible at this distance $r_c$. The actual radius of the circular orbit, as measured from the cone's symmetry axis,  is $\rho_c=r_c\sin\alpha$.  

Moreover, equation (\ref{10}) can be directly integrated to yield the general solution of the problem as

\begin{equation}\label{12}
t=\pm\int_{r_1}^r{dr'\over\sqrt{{2\over M}[{\cal E}-({A\over r'^2}+{B\over r'}+Cr')]}},
\end{equation}

\noindent where we have chosen the reference point as one of the turning points of the motion. Taking into account that $\dot r={(dr/ dt)}\dot\vartheta$, we  obtain, from equations (\ref{8}) and (\ref{10}), the orbit equation as
 
\begin{equation}\label{13}
\vartheta=\pm{1\over\sin^2\alpha\sqrt{2M}}\int_{r_1}^r{L_{z}dr'\over r'^2\sqrt{{\cal E}-
({A\over r'^2}+{B\over r'}+Cr')}},
\end{equation}

\noindent where ${r_1}$ corresponds to the  first turning point and the double sign corresponds to the clockwise and counterclockwise rotations, respectively. Equations (\ref{12}) and (\ref{13}) are the general form of the solution of the problem in terms of quadratures.  The integrals (\ref{13}) and (\ref{12}) can be expressed in terms of elliptic integrals of the first and of the second kind, but such expressions do not offer much insight into the properties of the motion. 
However, a simple analysis of figure 2 says that the motion is always bounded  between $r_1$ and $r_2$, and that, for given  $E$, $h$, and $L_z$, there can be only one circular orbit which  besides is stable. 
To determine the radius, $r_c$, of this orbit we must take $\dot u=0$ ---which means that the acceleration, and thus the force, in the $r$-direction vanishes---
in (\ref{1}) to get
\begin{equation}\label{19}
7 \dot\vartheta_c^2 r_c \sin^2\alpha- \left[{5g}  -  {2}  a \omega_c \dot\vartheta_c \right]\cos\alpha=0, 
\end{equation}

\noindent the suffix $c$ meaning magnitudes evaluated on the parallel circle $r=r_c$. Since the first term in equation (\ref{19}) is always positive and  $\cos\alpha>0$ because $0<\alpha<\pi/2$, the circular orbit  is only possible when the initial conditions are such that 

\begin{equation}\label{20}
\dot\vartheta_c\omega_c<{5g\over 2a}.
\end{equation}

\ni Equation (\ref{19}) is quite useful since it allows evaluating any one of the three quantities $r_c$, $\dot\vartheta_c$, or $\omega_c$ as a function of initial conditions of the remaining two for starting the motion on the circular orbit. If we know the initial values $\omega_0$ and $\dot\vartheta_0$, we could ascertain whether equation (\ref{19}) has a solution  or not, just by checking if those initial values satisfy equation (\ref{20}). Clearly equation (\ref{20}) holds if  $\omega_0$ and $\dot\vartheta_0$ have contrary signs independent of their magnitudes. Now equation (\ref{19}) has one solution for  $\omega_c$ with initial conditions $r_0$ and $\dot\vartheta_0$ while it has two solutions with initial conditions $r_0$ and $\omega_0$ one of opposite sign to  $\omega_0$ and another of the same sign, but satisfying equation (\ref{20}). 

Equation (\ref{19}) in terms of the constants of motion $L_z$ and $h$ is

\begin{equation}\label{21}
 5\,r_c^3\,M^2g \sin^2\alpha \cos\alpha- 2\, r_c\, ML_zh \sin^2\alpha\cos\alpha-7L_z^2\left( 1-\frac{5} {7} \cos^2\alpha\right)=0. 
\end{equation}

\ni This is a cubic equation for $r_c$ that  has always a real positive root; but let us pinpoint that, due to condition (\ref{20}), this equation is only valid when $ 2hL_z<5Mr_c^2g\sin^2\alpha$. The existence of only one real and positive root, follows from the fact that the last term in (\ref{21}) is less than zero, whereas the coefficient of the first term is positive \cite{uspensky}. However, it can be much simpler to just take a look at  figure 2, as it exhibits that there is one and only one circular orbit on the surface. \par

There is a further point worth mentioning, the existence of a relationship between the radius of the circular orbit and the $r$-values of the turning points, $r_i,\;i=1,2$, namely

\begin{equation}\label{global}
r_c^3=\frac{2 r_1^2 r_2^2} {r_1+r_2} + \frac {B} {C} \left(r_c -\frac {r_1r_2} {r_1 + r_2} \right).
\end{equation}

\noindent This is the sort of property called {\sl universal} in \cite{lopez02}; in fact, (\ref{global}) is very easily shown to reduce to  equation (13) of reference \cite{lopez02} in the point particle limit (that is, when $B=0$).

\section {Analogy with an Non Homogeneous Harmonic Oscillator  and the Orbit on a Freelly Falling Cone.}

  Let us substitute equations (\ref{6}) in the first of equations (\ref{1}), to get 

\begin{equation}\label{15}
\ddot{r}-r \dot{\varphi}={2\over 7}{\dot{\varphi}\over\sin\alpha}(a\omega)\cos\alpha-{5\over 7}g\cos\alpha,
\end{equation}

\noindent where  $\varphi\equiv\vartheta\sin\alpha$.    Multiplying by  $M$ and using equations(\ref{3}) and  (\ref{4}) in the right hand side of  equation (\ref{15}), we get

\begin{equation}\label{16}
M(\ddot r-r\dot\varphi^2)=f(r),
\end{equation}

\noindent where

\begin{equation}\label{16b}
f(r)\equiv {B'\over r^2}+{A^\prime\over r^3}-{5\over 7}Mg\cos\alpha,
\end{equation}

\noindent and

\begin{equation}\label{17}
\eqalign{ B'&\equiv{2\over 7}Lh\cot\alpha\cr 
A^\prime&\equiv{2\over 7}{L^2\cot^2\alpha\over M}\cr
L&\equiv M r^2\dot \varphi ={L_z\over \sin \alpha};\cr}
\end{equation}

\noindent therefore the equations of motion of the sphere are found, as we did in section 4,   analogous to the equations of motion of a particle in a central force field. The variables $r(t)$ and $\varphi(t)$ can be considered as the polar coordinates of a particle moving on a plane under the action of the ``central'' force $f(r)$ given in equation (\ref{16}). 
If, additionally we assume the cone is in free fall, that is, that $g=0$, then the orbit of the center of the sphere, $r=r(\varphi)$, can be easily obtained by rewriting  equation (\ref{15}) in terms of  $\varphi$,  using 
the relations 

\begin{equation}\label{17b}
{d\over dt}=\dot\varphi{d\over d\varphi}={L\over Mr^2}{d\over d\varphi}
\end{equation}

\noindent and

\begin{equation}\label{17c}
{d^2\over dt^2}={L^2\over Mr^2}{d\over d\varphi}\left({1\over r^2}{d\over d\varphi}\right).
\end{equation}

 \noindent Now,  introducing Binet's variable $W \equiv {1/ r}$,  the equation of motion  (\ref{16}) becomes, with $g=0$ and using (\ref{17b}) and (\ref{17c}),
 
\begin{equation}\label{17d}
{d^2 W\over d\varphi^2}+\left(1+{2\over 7}\cot^2\alpha\right)W=-{2\over 7} {Mh\cot\alpha \over L}
\end{equation}

\noindent or, in terms of the angle in the polar plane, $\vartheta={\varphi/ \sin\alpha}$, and the vertical component of the angular momentum, $L_z$, 

\begin{equation}\label{17e}
{d^2 W\over d\vartheta^2}+\left(1-{5\over 7}\cos^2\alpha\right)W=-{2\over 7} {Mh\cos\alpha\sin^2\alpha\over
L_z}.
\end{equation}

\noindent This is the well-known inhomogeneous differential equation for  harmonic motion.  If we define $\Omega^2\equiv \left(1-(5/ 7)\cos^2\alpha\right) $, the general solution can be written   

\begin{equation}\label{17f}
W\equiv {1\over r}={\cal A}\cos\left[\Omega (\vartheta-\vartheta_0)\right]- {2Mh\sin^2\alpha\cos\alpha\over 7L_z\Omega^2}.  
\end{equation}

\noindent The constants of integration  ${\cal A}$ and $\vartheta_0$ can be determined by the initial conditions. 

To clasify the solutions of the problem in free fall, we  recognize two cases,
\begin{description}
 \item{A)} when $h$ and  $L_z$ have different signs,  i.e. $hL_z<0$; or 
\item{B)} when $h$ and $L_z$ have the same sign, i.e., $hL_z>0$. 
\end{description}

\noindent In any case,  the general solution can be written in the form
 
\begin{equation}\label{17g}
W\equiv {1\over r}={\cos\alpha\over p} e\cos\left[\Omega(\vartheta-\vartheta_0)\pm 1\right]
\end{equation}

\ni with 
\begin{equation}\label{p}
p\equiv \frac {7|L_z| \Omega^2} {2M|h|\sin^2\alpha}
\end{equation}

\noindent or in the form

\begin{equation}\label{17h}
r(\vartheta)=r_{min}{e\pm 1\over e\cos\Omega(\vartheta-\vartheta_0)\pm 1}
\end{equation}

\noindent with

\begin{equation}\label{18}
r_{min}={p \over \cos\alpha(e\pm 1)},
\end{equation}

\noindent where $e$  and $\vartheta_0$ are  constants of integration. The sign in equation (\ref{18}) is  plus if  $hL_z<0$ and minus if  $hL_z>0$. Without loss of generality we can assume that $e\geq 0$, since its sign can be changed by substituting $\vartheta\rightarrow\vartheta+\pi$, that is, just by changing the orientation of the coordinates.

 The possible motions are best analysed case by case, as follows.

\begin{itemize}
\item [A)] In the case that  $hL_z>0$, we have the minus sign in  equation (\ref{18})  and therefore  $e>1$. In this case there are  two values of $\vartheta$ for which $e\cos\Omega(\vartheta-\vartheta_0)=1$. When  $\vartheta$ approaches any of these two values $r\rightarrow\infty$; therefore they define two asymptotes and the orbit corresponds  to a branch of an hyperbola. 

\item[B)] In the case that  $hL_z<0$, the plus sign is in order in  equation (\ref{18}). We find now four possibilities, $e=0$, $0<e<1$, $e=1$ and  $e>1$. The motion in these cases is as we describe in what follows.

\begin{description}

\item {B1.} If $e=0$, the orbit is the circle
 
\begin{equation}\label{18b}
r(\vartheta)=r_c={p\csc\alpha\left(1-{5\over 7}\cos^2\alpha\right) };
\end{equation}

\ni this relation coincides with (\ref{21}) if $g=0$.

\item{B2.} If  $0<e<1$, the orbit is confined to the strip  defined by the two parallel circles on the cone, having radii

\begin{equation}\label{18c}
r=r_{min}
\end{equation}

\noindent and

\begin{equation}\label{18d}
r=r_{max}=r_{min}{1+e\over 1-e}.
\end{equation}

The CM of the sphere touches the circle with radius $r=r_{max}$ when  

\begin{equation}\label{18e}
\vartheta-\vartheta_0={\pi\over \sqrt{1-{5\over 7}\cos^2\alpha}},
\end{equation}

\noindent which means that the apsidal angle $\psi$ (the angle swept by the CM's radius vector in going from $r_{min}$ to $r_{max}$) is 

\begin{equation}\label{18f}
\psi={\pi\over \sqrt{1-{5\over 7}\cos^2\alpha}}> \pi,
\end{equation}

\noindent thus the apsides advance in each period of rotation. As the orbit is symmetric with respect to the polar axis ($r(-\vartheta)=r(\vartheta)$), the angular displacement of the apsides in each rotation of the center of the sphere around the $z$-axis is 

\begin{equation}\label{18h}
\Delta\psi=2\pi[(1-{5\over 7}\cos^2\alpha)^{-1/2}-1]
\end{equation}

\noindent and  the angular velocity of precession  of the apsides is 

\begin{equation}\label{18i}
{\Delta\psi\over T}={2\pi\over T}[(1-{5\over 7}\cos^2\alpha)^{-1/2}-1]=\langle\dot \vartheta\rangle [(1-{5\over 7}\cos^2\alpha)^{-1/2}-1]
\end{equation}

\noindent where $T$ is the period of rotation of the CM around the $z$ axis, and

\begin{equation}\label{18j}
\langle \dot \vartheta\rangle \equiv {1\over T}\int_0^T\dot\vartheta dt\equiv {2\pi\over T}
\end{equation}

\ni is the mean value of $\dot\vartheta$. 

Since $(1-{5\cos^2\alpha}/7)^{1/2}$ can be shown to be an irrational number for $0<\alpha<\pi/2$, the orbit never repeats itself thus filling completely the strip that contains the orbit as $t \to \infty$. We may say that the orbit is {\sl ergodic} on the strip, that is, that eventually the orbit would densely fill all of the strip's surface. \par

\item {B3.} If $e=1$, we can easily verify that $r \to \infty$ as 
 
\begin{equation}\label{18k}
\vartheta\rightarrow\pm{\pi\over \sqrt{1-{5\over 7}\cos^2\alpha}}+\vartheta_0.
\end{equation}

\ni Therefore, the orbit is a branch of an hyperbola with asymptotes defined by 
  
\begin{equation}\label{18l}
\vartheta=\pm{\pi\over \sqrt{1-{5\over 7}\cos^2\alpha}}+\vartheta_0.
\end{equation}\par

\item {B4.} If  $e>1$, then there exist two values of  $\vartheta$ for which 
 
\begin{equation}\label{18m}
e\cos \sqrt{1-{5\over 7}\cos^2\alpha}(\vartheta-\vartheta_0)=-1,
\end{equation}

\noindent and therefore $r\to \infty$ when $\vartheta$ approaches any of these values, which simply define  the asymptotes of the corresponding branch of the hyperbola.\par

\end{description}
\end{itemize}

The only thing that remains to be done is to obtain the value of $e$ in terms of the constants of motion. For this we have only to susbstitute the values of  $r_{min}$ and of $p$ into equation (\ref{10}) (with $C=0$)  to get 

\begin{equation}\label{18n}
e=\sqrt{1+{49(1-{5\over 7}\cos^2\alpha){\cal E}\over 2Mh^2\cos^2\alpha\;}}.
\end{equation}

\noindent Note that $e$ depends only on ${\cal E}$ and $h$, and, furthermore, that $0<e<1$, if  ${\cal E}<0$; that $e=1$, if  ${\cal E}=0$; and that  $e>1$, if  ${\cal E}>0$. But, if $hL_z>0$ then $B>0$ and therefore ${\cal E}>0$, from which we have  $e>1$ always, consequently this case is equivalent to motion in a repulsive force field.\par

\section {Conclusions}
              We have studied the motion of a sphere rolling  on the inner surface of a right circular cone with opening angle $\alpha$ under the assumption that it does not slip. The motion was first considered under the action of gravity and we managed to obtain an implicit solution in quadratures. We  established the existence of bounded motions and of circular orbits in the system analysing an effective potential in which the equivalent single particle system moves. With the help of this effective potential we were able to reduce the problem to quadratures obtaining implicitly $r(t)$ and thus $\theta(t)$. 

However,  to actually integrate the equations of motion for obtaining the centre's of mass orbit in terms of elementary functions,  we required changing to a freely falling frame. In such a  frame we have been able to obtain solutions for the CM orbit analogous to ones in a related central problem. We have been able to show, with the help of this relation, that the CM can move on hyperbolic or  on preceding elliptic,  but {\sl never} on parabolic orbits. 

The analogy between motions in our  system  with central-field motion and specially with an inhomogeneous oscillator in the freely falling case, should be pointed out as interesting and quite amusing. This aspect of the behaviour has  a certain pedagogical significance as can serve to emphasise that the same equations have always the same solutions; it does not really matter that we are describing different physical systems.

\ack ALSB has been partially supported by a PAPIIT-UNAM grant (108302), he expresses his thaks to the Computational Science Research Center of SDSU and particularly to Jose E Castillo and Ricardo Carretero-Gonz\'alez for computational support. We acknowledge with thanks the help of H. N. N\'u\~nez-Y\'epez. The cheerful collaboration of A Simon and the late M Nick of  Rolando California, and of H Kranken,  P M Lobitta,  D Yoli, V Binny of the Monte Bello gang, is also gratefully acknowledged. This paper is dedicated to the memory of our beloved friends P M Botitas, F C Sadi, M K Dochi, and  C Suri.


\vskip 40pt

\begin{thebibliography}{222}

%
\bibitem{ll} Landau L and Lifshitz E M 1976 {\sl Mechanics} (Oxford: Pergamon) Ch VI 3rd edition

\bibitem{sommerfeld} Sommerfeld A 1950 {\sl Mechanics} (New York: Academic Press) Ch IV
ez/Y-epez, 
\bibitem{goldstein} Goldstein H  1950  {\sl Classical Mechanics} (New York:  Addison-Wesley) 

\bibitem{flores72} Flores J, Del Rio A G, Calles A, Riveros H 1972 Am.\ J. Phys.\ {\bf 40} 595

\bibitem{soodak02} Soodak H 2002  Am.\ J. Phys.\  {\bf 70} 815

\bibitem{lopez02}  L\'opez-Ruiz R and Pacheco A F 2002   Eur.\ J. Phys.\ {\bf 23} 579 
 
\bibitem{96} C. M. Arizmendi, R. Carretero-Gonz\'alez, H. N. N\'u\~nez-Y\'epez, A. L. Salas-Brito, {\it  The curvature criterion and the dynamics of a rolling elastic cylinder},  in {\it   New Trends in HS\&CM: Advanced Series in Nonlinear Dynamics} {Vol.\ 8}, Eds.\ E.\ Lacomba and J.\ Llibre, World Scientific, (1996) 1--13.  

\bibitem{theron02} Theron W F D 2000 Am.\ J. Phys.\ {\bf 68} 812 

\bibitem{carnero97} Carnero C, Carpena P and Aguiar J 1997 Eur.\ J. Phys.\ {\bf 18} 409

\bibitem{previous} Fern\'andez-Chapou J L 1998 in {\sl Geometric Control and Non-Holonomic Mechanics} Jurdjevic V and Sharpe R W eds (Providence: American Mathematical Society and Canadian Mathematical Society) 199

\bibitem{efroimsky00} Efroimsky M 2000 J. Math.\ Phys.\  {\bf 41} 1854

\bibitem{macie03} Maciejewski A J, Przybylska 2003 Nonintegrability of the problem of a small satellite in gravitational and magnetic fields arXiv math-ph/0308010

\bibitem{ranada95} Barrientos M, P\'erez A and Ra\~nada A F 1995 Eur.\ J. Phys.\ {\bf 16} 106

\bibitem{fajans00} Fajans J 2000 Am.\ J. Phys.\ {\bf 68} 654

\bibitem{uspensky} Uspensky J V 1997 {\sl Teor\'{\i}a de Ecuaciones} (Mexico City: Limusa) Ch V

\end{thebibliography}
\end{document}